%% file: main.tex
\documentclass[letterpaper, 10 pt, conference]{ieeeconf}  

\IEEEoverridecommandlockouts                              

\usepackage{url}
\usepackage{cite}
\usepackage{amsmath,amssymb,amsfonts}
\usepackage{algorithmic}
\usepackage{graphicx}
\usepackage{textcomp}
\usepackage[dvipsnames]{xcolor}
\usepackage{soul}
\usepackage{multirow,multicol}
\usepackage{tikz}
\usepackage[linesnumbered,ruled,vlined]{algorithm2e}
\SetKwInput{KwInput}{Input}                
\SetKwInput{KwOutput}{Output}

\usetikzlibrary{patterns,decorations.pathmorphing,positioning}

\def\BibTeX{{\rm B\kern-.05em{\sc i\kern-.025em b}\kern-.08em
    T\kern-.1667em\lower.7ex\hbox{E}\kern-.125emX}}

\input{macros.tex}

\newcommand{\response}[1]{\textcolor{black}{#1}}
\newcommand{\responseOne}[1]{\textcolor{black}{#1}}

\newcommand{\responseThree}[1]{\textcolor{black}{#1}}
\newcommand{\responseFive}[1]{\textcolor{black}{#1}}

\title{\LARGE \bf Sensor Placement with Optimal Precision for Temperature Estimation of Battery Systems}

\author{Vedang M. Deshpande$^{1}$, Raktim Bhattacharya$^{2}$, Kamesh Subbarao$^{3}$ 
\thanks{The first author gratefully acknowledges support from the PhD Graduate Excellence Fellowship administered by the Department of Aerospace Engineering, Texas A\&M University.}
\thanks{$^{1}$Vedang M. Deshpande is a Ph.D. student in Aerospace Engineering, Texas A\&M University, College Station, TX 77843, USA. {\tt\small vedang.deshpande@tamu.edu}}
\thanks{$^{2}$Raktim Bhattacharya is Associate Professor in Aerospace Engineering,
Electrical \& Computer Engineering, Texas A\&M University, College Station, TX 77843, USA. {\tt\small raktim@tamu.edu}}
\thanks{$^{3}$Kamesh Subbarao is Professor in Mechanical and Aerospace Engineering Department at The University of Texas at Arlington, Arlington, TX 76019, USA. {\tt\small subbarao@uta.edu}}%
}

\begin{document}
\maketitle
\thispagestyle{empty} 
\begin{abstract}
The temperature distribution in the battery significantly impacts the short-term and long-term performance of battery systems. Therefore, efficient, safe, and reliable battery system operation requires an accurate estimation of the temperature field. The current industry standard for sensors to battery cell ratio is quite frugal. Thus, the problem of sensor placement for accurate temperature estimation becomes non-trivial, especially for large-scale systems. In this paper, we explore a greedy approach for sensor placement suitable for large-scale battery systems. An observer to estimate the thermal field is designed in an $\Hinf$ framework while simultaneously minimizing the sensor precisions, thus lowering the overall thermal management system's economic cost.
\end{abstract}

\section{Introduction}
Batteries are used for energy storage in various applications, including portable consumer electronics, electric ground, aerospace vehicles, and recent push towards large-scale electric grid energy storage systems. Irrespective of the application, efficient, safe, and reliable operation of a battery system requires monitoring and regulating key battery states, e.g., temperature, power, charge, and health, collectively known as battery management \cite{lin_modeling_2019, hu_state_2019}.

Improper operating temperatures and temperature gradients within the battery severely impact both short-term performances (power capability, efficiency, and self-discharge rate) and long-term performance (power fade, capacity degradation)  \cite{bandhauer_critical_2011}. Notably, among the widely used Li-Ion batteries, a large amount of heat is typically accumulated during the charging and discharging process \cite{li-ion-Review, heatRate}. Further, the accumulated heat doesn't dissipate evenly or at predictable rates due to the Li-Ion battery structure, which seriously degrades the performance and the batteries' useful life. Thus, thermal management based on the estimation of temperatures is a critical aspect of battery management.

Most engineering battery packs consist of several battery cells packed in strings, and the number of cells in the pack can range from few tens to several thousand \cite{eberhard_bit_2006}.
The surface temperature of battery cells is one of the few quantities that can be measured by typically available sensors in the system \cite{lu_review_2013}. However, present average sensors-to-cells ratio in the industry is about 1 thermal sensor for every 10 cells \cite{lin_robust_2020, samad_improved_2016}. Therefore, the problem of sensor placement at design-time, i.e., selecting optimal locations in a string of battery cells to place thermal sensors, becomes non-trivial, especially for large-scale systems consisting of thousands of cells.

The problem of sensor placement for maximizing a certain measure of observability of thermal estimation system was studied in \cite{lin_parameterization_2013, wolf_optimizing_2012, lystianingrum_observability_2014}. Sensor selection problem to achieve the best estimator performance in terms of $\Hinf$ norm of the error system in the presence of model uncertainty was discussed in \cite{lin_robust_2020}. In \cite{lin_parameterization_2013, lin_robust_2020, lystianingrum_observability_2014}, the optimal locations are determined by exhaustive search, i.e. solving the optimization problem for every possible combination, which becomes intractable for systems of even moderate sizes. An ad-hoc approach to avoid computational burden for systems with a large number of cells is proposed in \cite{samad_improved_2016} to divide the string into multiple sections of a smaller number of cells, and the sensor placement problem was solved locally in each section.  A modified greedy approach involving eigenanalysis of the governing partial differential heat equation to select sensors in a tractable way was discussed in \cite{wolf_optimizing_2012}.  

The afore discussed works do not take into account the sensor precisions while selecting the optimal sensor locations. In general, sensors with high precision cost more. Therefore, to lower the overall economic cost, both the number and the sensors' precisions should be minimized.
This paper presents a framework for sensor location selection while maximizing the allowable sensor noise (or \responseFive{minimizing} sensor precision).
Thus, the proposed framework can be used for designing a temperature estimator anticipating the degradation of the sensor over the lifespan of the battery.

The objective of this work is threefold: (i) select a specified number of locations to place temperature sensors in a string of battery cells, (ii) minimize the required sensor precision, and (iii) design an observer to estimate the thermal field using the selected sensors such that the estimation errors are bounded. This is the first time sensor placement problem with optimal precisions has been studied for large-scale battery systems to the best of our knowledge. The optimal observer design problem is posed in an $\Hinf$ formulation, and a greedy algorithm is discussed as a tractable approach to solve the sensor selection problem.

\subsubsection*{Organization} The thermal model of battery systems is discussed briefly in Section \ref{sec:batMod}. Problem formulation for the observer design and the greedy algorithm are presented in Section \ref{sec:prob}. Section \ref{sec:res} presents some numerical results followed by the concluding remarks in Section \ref{sec:concl}.

\subsubsection*{Notation}
The set of real numbers is denoted by $\Real$. Bold uppercase (lowercase) letters denote matrices (column vectors). An identity matrix and a zero matrix of suitable dimensions are denoted by $\I{}$ and $\vo{0}$ respectively. Symmetric positive (negative) definite matrices are denoted by $\X>0$ ($\X<0$). A vector's exponent is interpreted elementwise.

\section{Battery Model for Temperature Estimation} \label{sec:batMod}
\begin{figure}
    \centering
    \includegraphics[trim={0cm 0.25cm 0cm 0cm},clip,width=0.38\textwidth]{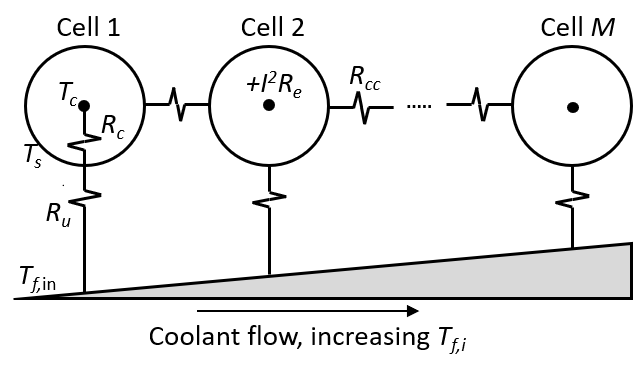}
    \caption{Thermal model of battery cells packed in a string}
    \label{fig:string}
    \vspace{-16pt}
\end{figure}
Let us consider a battery pack consisting of $\nbat$ cells packed in a string as shown in \fig{string}. We use the thermal model developed in \cite{lin_parameterization_2013} for such battery systems. This model is briefly discussed below. A detailed discussion on the model can be found in  \cite{lin_parameterization_2013, lin_robust_2020} and the references therein.

The temperature field of the \ith cell in the string is characterized by the core temperature $T_{c,i}$ and the surface temperature $T_{s,i}$. Thermal dynamics is given by
\begin{subequations}
  \begin{align}
C_{c,i} \frac{d T_{c,i}}{dt} & = \bar{I}^2 R_{e,i} + \frac{T_{s,i}-T_{c,i}}{R_{c,i}}, \eqnlabel{thermal_dyn_a} \\
C_{s,i} \frac{d T_{s,i}}{dt} & = \frac{T_{f,i}-T_{s,i}}{R_{u,i}} - \frac{T_{s,i}-T_{c,i}}{R_{c,i}} + Q_{cc,i}, \eqnlabel{thermal_dyn_b}
\end{align}  \eqnlabel{thermal_dyn}
\end{subequations}
where $C_{c,i}$ and $C_{s,i}$ denote heat capacities of core and surface of the \ith cell, respectively. \response{Equation \eqn{thermal_dyn_a} is the energy conservation or heat balance equation for a cell's core. The left side of  \eqn{thermal_dyn_a} denotes the rate of change of internal heat energy of the core (proportional to the rate of change of core temperature), and it is equal to the sum of rates of heat generation within the cell and heat exchange with the surface. The heat is generated due to Joule heating, where $R_{e,i}$ is the internal resistance of the cell and $\bar{I}$ is the current modeled as an input. The heat exchange rate with the surface is modeled as the heat conduction due to temperature difference over a lumped thermal resistance $R_{c,i}$.}

\response{Similarly, the heat balance equation for a cell's surface is given in \eqn{thermal_dyn_b}.} 
We denote by $T_{f,i}$ the temperature of coolant at the \ith cell. The convection thermal resistance between the surface of \ith cell and the coolant is denoted by $R_{u,i}$. The coolant temperature at the \ith cell is then obtained by balancing the heat flow over the preceding cell. \response{In particular, the difference between $T_{f,i}$ and $T_{f,i-1}$ is obtained by dividing the heat withdrawn from the $(i-1)^{\text{th}}$ cell due to convection by the heat capacity rate of coolant, $\bar{C}_f$, i.e.} 
\begin{align*}
     T_{f,i} =  T_{f,i-1} + \frac{T_{s,i-1} - T_{f,i-1}}{\bar{C}_f R_{u,i}}, \quad  2\leq i\leq\nbat,
  \end{align*}
and $T_{f,1}=:T_{f,\text{in}}$, where the inlet coolant temperature $T_{f,\text{in}}$ is assumed to be known and will be modeled as an input. 

The heat conduction between two consecutive cells is denoted by $Q_{cc,i}$ and is modeled as a heat flow over the lumped conduction resistance $R_{cc}$
\begin{equation*} Q_{cc,i} = \left\{
  \begin{aligned}
     & \frac{T_{s,2} - T_{s,1}}{R_{cc}}, & i=1,  \\
     & \frac{T_{s,i-1} - 2T_{s,i} + T_{s,i+1}}{R_{cc}}, \  & 2\leq i\leq\nbat-1, \\
     & \frac{T_{s,\nbat-1} - T_{s,\nbat}}{R_{cc}}, & i=\nbat.
  \end{aligned} \right.
\end{equation*}
Thermal dynamics \eqn{thermal_dyn} can be written as  the following linear time invariant system.
\begin{align}
  \xdot(t) = \A\x(t) + \Bu\u(t) + \Bd d(t) ,\eqnlabel{sys}
\end{align}
where  $\x\in\Real^{2\nbat}$, is the state vector and $\u\in\Real^{2}$ is the input vector defined as follows
\begin{align*}
  \x &:=  \begin{bmatrix} T_{c,1} & T_{s,1} & T_{c,2} & T_{s,2} & \cdots & T_{c,\nbat} & T_{s,\nbat} \end{bmatrix}^T, \\
  \u &:= \begin{bmatrix} \bar{I}^2 & T_{f,\text{in}} \end{bmatrix}^T.
\end{align*}
The system matrices are $\A$ and $\Bu$ are given in \eqn{sys_mat}. We assume that all cells are identical, therefore, the subscript $i$ is omitted in \eqn{sys_mat} for simplifying the notations.
\begin{table*}[ht!]
  \begin{equation}
    \begin{aligned}
    \A  &= \begin{bmatrix}
              \frac{-1}{C_{c} R_{c}} & \frac{1}{C_{c} R_{c}} & 0 & 0 & 0 & 0 & \cdots & 0 & 0 & 0 \\
              \frac{1}{C_{s} R_{c}} & -V & 0 & \frac{1}{C_{s} R_{cc}} & 0 & 0 & \cdots & 0 & 0 & 0 \\
              0 & 0 & \frac{-1}{C_{c} R_{c}} & \frac{1}{C_{c} R_{c}} & 0 & 0 & \cdots & 0 & 0 & 0 \\
              0 & \frac{1}{C_{s}} \left(\frac{1}{\bar{C}_f R_u^2} + \frac{1}{R_{cc}}\right) & \frac{1}{C_{s} R_{c}} &  \frac{-1}{C_{s}} \left(\frac{1}{R_{c}} + \frac{1}{R_u} + \frac{2}{R_{cc}}\right) & 0 & \frac{1}{C_{s} R_{cc}} & \cdots & 0 & 0 & 0 \\
              & & & & & & \ddots & & & \\
              0 & 0 & 0 & 0 & 0 & 0 & \cdots & 0 & \frac{-1}{C_{c} R_{c}} & \frac{1}{C_{c} R_{c}} \\
              0 & \frac{\left(1-\frac{1}{\bar{C}_f R_u}\right)^{\nbat-2} }{C_{s} \bar{C}_f R_u^2} & 0 & \frac{\left(1-\frac{1}{\bar{C}_f R_u}\right)^{\nbat-3} }{C_{s} \bar{C}_f R_u^2} & 0 & \cdots & \cdots & \frac{1}{C_{s}} \left(\frac{1}{\bar{C}_f R_u^2} + \frac{1}{R_{cc}}\right) & \frac{1}{C_{s} R_{c}}  & -V
      \end{bmatrix}, \\
    V &:= \frac{1}{C_{s}}  \left(\frac{1}{R_{c}} + \frac{1}{R_u} + \frac{1}{R_{cc}}\right), \qquad
    \Bu = \begin{bmatrix}
              \frac{R_{e}}{C_{c}} & 0 & \frac{R_{e}}{C_{c}} & 0 & \cdots & \frac{R_{e}}{C_{c}} & 0 \\
              0 & \frac{1}{C_{s}R_u} & 0 & \frac{1}{C_{s}R_u}\left(1-\frac{1}{\bar{C}_f R_u}\right) & \cdots & 0 & \frac{1}{C_{s}R_u}\left(1-\frac{1}{\bar{C}_f R_u}\right)^{\nbat-1}
          \end{bmatrix}^T. \\
    \end{aligned} \eqnlabel{sys_mat}
  \end{equation}
  \vspace{-0.6cm}
\end{table*}
The process noise is denoted by $d(t)$ that could account for several disturbances such as unintended fluctuations in the inputs to the system or unmodeled physical processes in the battery pack, significantly affecting the thermal dynamics. In this paper, we assume that a scalar disturbance signal $d\in\Real$ enters the system via $T_{f,\text{in}}$ input channel. Therefore, the matrix $\Bd$ is given by
\begin{align*}
      \Bd = \Bu \begin{bmatrix} 0 & S_d \end{bmatrix}^T,
\end{align*}
where $S_d > 0$ is a known scalar used for normalization of the disturbance signal $d(t)$. Similarly, other disturbances could also be incorporated into the system equations.

The parameter values for $A123\ 26650\ \text{LiFePO}_4$ battery cells used for numerical simulations are taken from \cite{lin_robust_2020} and tabulated in \eqn{param_vals}.
\begin{align}
    & C_c = 67 J/K, \,  C_s = 4.5 J/K, \, \bar{C}_f = 2.6 J/Ks, \,  R_e = 0.01 \Omega,  \nonumber \\
    & R_c = 1.83 K/W, \,  R_u = 5 K/W, \, R_{cc} = 0.2 K/W. \eqnlabel{param_vals}
\end{align}
The parameter values in \eqn{param_vals} are obtained by experiments and by calculations based on the battery's geometry. The actual battery model will suffer from uncertainties in these parameters, resulting in the estimation algorithm's poor performance. A sparse sensing framework incorporating polytopic uncertainties in the battery model is discussed in \cite{lin_robust_2020}. A generalized framework for sparse sensing with optimal precision in the presence of structured and unstructured uncertainty is presented in \cite{deshpande_sparseRobHinf_ACC2021}, which can be used for thermal estimation of uncertain battery systems. However, for this short paper, we work with the deterministic battery model given by \eqn{sys}.

We assume that the available set of sensors measure the surface temperature of each cell, i.e. $T_{s,i}$, $i=1,2\cdots M$. Therefore, the measurement equation for the sensor placed at the \ith cell is given by
\response{
\begin{align}
  y_i(t) = \vo{c}_i^T \x(t) + \sigma_i n_i(t), \eqnlabel{meas_i}
\end{align}
where $\vo{c}_i\in\Real^{2\nbat}$ is a vector} with $(2i)^\text{th}$ element equal to unity and all other elements zero.
The measurement at the \ith sensor location $y_i(t)$ is assumed to be corrupted by a noise signal $n_i(t)$ normalized by an unknown scalar $\sigma_i>0$.
In our case, the following terms: \ith sensor, \ith sensor location, or a sensor placed at the \ith cell have the same meaning and will be used interchangeably throughout the paper.
We assume that the process disturbance $d(t)$ and sensor noises $n_i(t)$ are \textit{power signals}. For a signal $n_i(t)$, let us define
\begin{align*}
   \pow{n_i}:= \left( \lim_{\tau\rightarrow \infty} \frac{1}{2\tau} \int_{-\tau}^{\tau} |n_i(t)|^2 dt \right)^{1/2},
\end{align*}
and the signal $n_i(t)$ is said to be a power signal if the limit under the square-root exists \cite{doyle_feedback_2009}.
Without loss of generality, it can be assumed that the signals $d(t)$ and $n_i(t)$ have unit power, i.e. $\pow{n_i} = 1$. Let us denote the effective noise signal as observed in the measurement $y_i(t)$ of the \ith sensor by $\tilde{n}_i(t)$, i.e. $ \tilde{n}_i(t) := \scln_in_i(t)$.

It immediately follows that $\pow{\tilde{n}_i} = \scln_i \pow{n_i} = \scln_i$.
We define the precision $\precs_i$ of the \ith sensor as the inverse of squared-power of effective noise in its measurements, i.e.
\begin{align}
    \precs_i := 1/\left(\pow{\tilde{n}_i}\right)^2 =  1/\scln_i^2. \eqnlabel{def_prec}
\end{align}

Precision of temperature sensors are sometimes also specified in terms of an upper bound on absolute value of the noise e.g. $|\tilde{n}_i(t)|\leq\scln_i$, or equivalently $\norm{\tilde{n}_i(t)}{\infty} := \sup_t|\tilde{n}_i(t)| = \scln_i$. In this case it is straightforward to show that $\pow{\tilde{n}_i}\leq\norm{\tilde{n}_i(t)}{\infty} = \scln_i$ \cite{doyle_feedback_2009}. Therefore, sensor precision alternatively can also be defined in terms of the $\infty$-norm of the noise signal.

Following the notation in \cite{deshpande_sparseH2Hinf_LCSS2021}, we define an output vector of interest, $\z(t)$, that we wish to estimate
\begin{align*}
  \z(t) = \Cz\x(t),
\end{align*}
where $\Cz$ is a real matrix of appropriate dimensions. Since our objective is to estimate the temperature field in the battery pack characterized by the full state vector $\x(t)$, we choose $\Cz$ to be an identity matrix, i.e., $\Cz = \I{}$. In a particular scenario where it is of interest to estimate temperatures only of a few selected cells or assign unequal importance to estimation errors in different temperature states, the matrix $\Cz$ can be selected accordingly.
Next, we discuss problem formulation for the sensor placement with optimal precisions.
\section{Problem Formulation} \label{sec:prob}
Sensor placement with optimal precision is a combinatorial problem and becomes intractable for large-scale systems. Therefore, we decompose the problem into two parts. First, we design an observer with optimal precision in Section \ref{sec:obs_des}, and for the observer design, we assume that a set of sensors is given.
Second, the problem of selecting a suitable set of sensors is discussed in Section \ref{sec:sens_plac} where we present a greedy algorithm to search over the possible sensor locations.

\subsection{Observer Design} \label{sec:obs_des}
Let $\Sens:=\{1,2\cdots\nbat\}$ denote the set of all cell indices at which a temperature sensor could be placed. Suppose we have been given a subset $\set{Q}=\{q_1,q_2\cdots q_{\ny}\}\subseteq\Sens$ of cardinality $|\set{Q}| = \ny$ such that the sensor measurement vector $\y\in\Real^{\ny}$ and the measurement matrix $\Cy\in\Real^{\ny\times 2\nbat}$ are given by
\response{
\begin{align*}
  \y  &:=  \begin{bmatrix} y_{q_1} & y_{q_2} & \cdots & y_{q_{\ny}} \end{bmatrix}^T, \\
 \Cy &:=  \begin{bmatrix} \vo{c}_{q_1} & \vo{c}_{q_2} & \cdots & \vo{c}_{q_{\ny}} \end{bmatrix}^T.
\end{align*}
}
Consequently, sensor measurement equations \eqn{meas_i} are written in a compact form as follows
\begin{align}
    \y(t) &= \Cy\x(t) +  \diag(\vo{\scln})\n(t), \eqnlabel{sys_meas}
\end{align}
where the vectors $\vo{\scln},\n\in\Real^{\ny}$ are defined as $\vo{\scln}:=[\scln_{q_1} ,\scln_{q_2}\cdots \scln_{q_{\ny}}]^T$, $\n:=[n_{q_1},n_{q_2}\cdots n_{q_{\ny}}]^T$, and $\diag(\vo{\scln})$ denotes a diagonal matrix with the vector $\vo{\scln}$ as its principal diagonal. In accordance with \eqn{def_prec}, let us define the precision vector \response{ $\vo{\precs}:=\vo{\scln}^{-2}$.}

The goal is to design an observer with bounded errors and simultaneously minimize the sensor precisions. We consider a Luenberger observer of the following form
\begin{subequations}
\begin{align}
    \dot{\hat{\x}}(t) =& \A\hat{\x}(t) + \Bu\u(t) + \Lg(\hat{\y}(t)-\y(t)) \eqnlabel{obs_1}\\
    \hat{\z}(t) =& \Cz\hat{\x}(t)
\end{align} \eqnlabel{obs}
\end{subequations}
where $\hat{\x}(t)$ is an estimate of the state vector $\x(t)$, $\Lg\in\Real^{2\nbat\times\ny}$ is an unknown observer gain to be determined, and $\hat{\y}(t) := \Cy\hat{\x}(t)$. Let us define the state estimation error $\xerr(t)$ and the error in estimate of $\z(t)$ as
\begin{align*}
    \xerr(t) := \x(t)-\hat{\x}(t), \text{ and }
    \zerr(t) := \z(t)-\hat{\z}(t). 
\end{align*}
Subtracting \eqn{obs_1} from \eqn{sys}, we get the following error system
\begin{equation}
  \begin{aligned}
      \dot{\xerr}(t) &= \left(\A+\Lg\Cy\right)\xerr(t) + \left(\Bw+\Lg\Dw\right)\w(t),  \\
      \zerr(t) &= \Cz\xerr(t).
  \end{aligned} \eqnlabel{obs_err}
\end{equation}
where $\w(t) :=\begin{bmatrix}d(t) & \n^T(t)\end{bmatrix}^T$, $\Bw := \begin{bmatrix}\Bd & \vo{0}\end{bmatrix}$ and $\Dw := \begin{bmatrix}\vo{0} & \diag(\vo{\scln})\end{bmatrix}$.

\responseThree{The matrix $\left(\A+\Lg\Cy\right)$ must be Hurwitz to ensure stability of the error system \eqn{obs_err}. The existence of such a stabilizing observer gain $\Lg$ requires that the pair $\left(\Cy,\A\right)$ is detectable.}
Since the thermal dynamics \eqn{thermal_dyn} and hence the matrix $\A$ are stable, the detectability condition is always satisfied \cite{lin_robust_2020}.

It is desired that the effect of process and sensor noise on the estimation error is minimal or bounded. The effect of the external disturbances on the estimation error $\zerr(t)$ can be quantified by a suitable norm of the transfer function matrix $\GO$ from $\w(t)$ to $\zerr(t)$ of the system \eqn{obs_err}, given by
\begin{align}
    \GO := & \Cz\big(s\I{}-\A-\Lg\Cy\big)^{-1} \big(\Bw +\Lg\Dw\big), \eqnlabel{G_def}
\end{align}
where $s$ denotes the complex variable.

In this work, we use $\Hinf$ norm of the transfer function as a measure to bound the estimation error. The  $\Hinf$ norm of the transfer matrix \eqn{G_def} is defined as follows \cite{zhou_robust_1996}
\begin{align*}
    \norm{\GO}{\Hinf} := \sup_{\omega\in\Real} \bar{\sigma}_{\max}\left[\vo{\mathcal{G}}(j\omega)\right] ,
\end{align*}
where $\bar{\sigma}_{\max}$ denotes the maximum singular value of a matrix and $j$ is the unit imaginary number. $\Hinf$ norm serves as an important measure of the observer performance since it can be used to bound different measures of the estimation error $\zerr(t)$. The following upper bound on the estimation error can be specified in terms of $\Hinf$ norm \cite{doyle_feedback_2009, zhou_robust_1996}
\begin{align*}
  \pow{\zerr} \leq \norm{\GO}{\Hinf} \pow{\w}.
\end{align*}
A similar inequality also holds for energy or $\mathcal{L}_2$ norm of the disturbance and error signals.

The performance of the observer is specified by an upper bound on the $\Hinf$ norm of the observer error system  \eqn{obs_err}, i.e. $\norm{\GO}{\Hinf}<\gamma$ where  $\gamma >0$ is a given parameter.
Therefore, we seek to determine the sensor precisions $\vo{\precs}$ such that
$\norm{\GO}{\Hinf}<\gamma$, and $l_1$-norm of the precision vector $\norm{\vo{\precs}}{1}$ is minimized. To this end, we adopt the result from \cite{deshpande_sparseH2Hinf_LCSS2021} for $\Hinf$ observer design with the optimal precisions to write the inequality $\norm{\GO}{\Hinf}<\gamma$ as a matrix inequality.

In particular, the following holds \cite{deshpande_sparseH2Hinf_LCSS2021, deshpande_ADMMGreedy_TAC2021}
\begin{align*}
  \norm{\GO}{\Hinf}<\gamma \iff \exists \X>0 \text{ such that } \M(\vo{\precs},\X,\Y)  < 0
\end{align*}
where $\M(\vo{\precs},\X,\Y)$ is a matrix defined as follows
\begin{align*}
\M (\vo{\precs},\X,\Y) := \begin{bmatrix} \M_{11}  & \M_{12} & \Cz^T & \Y \\
               \ast & -\gamma\I{} & \vo{0}    & \vo{0} \\
                \ast &  \ast    & -\gamma\I{}     & \vo{0} \\
               \ast &  \ast    & \ast & -\gamma \ \diag(\vo{\precs}) \end{bmatrix}.
\end{align*}
The asterisks denote symmetric part of the matrix, and its component matrices are defined as $\Y:= \X\Lg$, $\M_{12} :=\X\Bd$, and $\M_{11} :=(\X\A+\Y\Cy) + (\X\A+\Y\Cy)^T$.
Therefore, the optimization problem is stated as follows
\begin{align}
\min\limits_{\vo{\precs}>0,\X>0,\Y}\quad \norm{\vo{\precs}}{1}
\text{\response{ subject to }} & \M(\vo{\precs},\X,\Y) < 0 .\eqnlabel{hinf_thm_obs}
\end{align}
If a feasible solution is found, then the observer gain is recovered as $\Lg = \X^{-1}\Y$ \cite{deshpande_sparseH2Hinf_LCSS2021}.

\begin{remark}
 The cost function in the optimization problem \eqn{hinf_thm_obs} can be trivially replaced with the \textit{weighted} $l_1$-norm of the sensor precisions \cite{deshpande_sparseH2Hinf_LCSS2021}. In a complex battery model where different kinds of sensors are used for battery health monitoring, weighted $l_1$-norm can penalize different sensor precisions to different extents accounting for the difference in sensors' economic cost. 
\end{remark}

 The convex optimization problem \eqn{hinf_thm_obs} is a semi-definite program (SDP) which can be solved using standard software packages such as \texttt{CVX} \cite{grant_cvx_2020}. However, general-purpose solvers do not scale well as the system dimension is increased. Customized optimization algorithms that exploit the problem's local structure could be employed to solve the precision minimization problem efficiently. An interested reader is referred to \cite{deshpande_ADMMGreedy_TAC2021} for an alternating direction method of multipliers (ADMM) based algorithm to solve \eqn{hinf_thm_obs} which scales relatively better for large-scale systems.

\subsection{Sensor Placement} \label{sec:sens_plac}
\begin{algorithm}[b!]
\SetAlgoLined
 \KwInput{The set of all sensors $\Sens$, performance index $\gamma$, the number of sensors to be selected $\ks$}
 \KwOutput{A set $\set{Q}$ of the selected sensors}
 Initialize: $\set{Q}\leftarrow \Sens$\\
 \For{$k=1,2 \cdots (\ns-\ks)$}{
    \For{$i\in\set{Q}$}{
    Define new set $\set{Q}_i:=\set{Q}\setminus \{i\}$ \\
    Solve \eqn{hinf_thm_obs} for the set $\set{Q}_i$ and store the optimal cost function value as $v_i$ \hfill {\footnotesize $\qquad\qquad\qquad$ // \texttt{Set $v_i$ to $+\infty$ if \eqn{hinf_thm_obs} is infeasible}}
  }
  Determine the set index with the least cost, i.e.
  $i^{\ast}:=\arg\min_{i} v_i \ $.  {\footnotesize{// \texttt{Break ties arbitrarily}}} \\
  \eIf{$v_{i^{\ast}}<\infty$}{
  Eliminate sensor $i^{\ast}$, i.e., update the set $\set{Q}\leftarrow \set{Q}_{i^{\ast}}$
  }{
  No feasible solution found. Return a null set $\set{Q}\leftarrow \emptyset$, and exit the algorithm.
  }
  }
 \caption{Greedy algorithm}
 \label{algo:GSE}
 \end{algorithm}
In most sensor placement problems, the selected set of sensors, $\set{Q}$, must satisfy a certain cardinality constraint. Therefore, we aim to select $\nsens$ out of total $\nbat$ battery cells for placing sensors that will measure their surface temperatures while simultaneously minimizing the cost function in \eqn{hinf_thm_obs}.

A brute force method to obtain the optimal solution is to perform an exhaustive search, i.e., to solve the optimization problem \eqn{hinf_thm_obs} for all possible combinations of $m$ cells. A similar approach was used in \cite{lin_robust_2020} to identity the best location to place a sensor in a string of 10 cells.
\responseThree{This approach, although guaranteed to yield the optimal solution, becomes intractable for large-scale battery systems as the number of possible combinations, $\frac{\nbat!}{\nsens!(\nbat-\nsens)!}$, grow with the number of cells in the battery.}

\responseThree{Algorithms based on greedy methods have become popular choice for obtaining a sub-optimal solution in a tractable way for the otherwise intractable problem of sensor placement, for instance, see \cite{shamaiah_greedy_2010,summers_submodularity_2016,zhang_sensor_2017}. In our recent work \cite{deshpande_ADMMGreedy_TAC2021}, we proposed a greedy method for efficiently selecting sensors while simultaneously minimizing the sensor precisions. Similar approach  for selecting cell locations (or sensors) in the battery system is outlined in Algorithm \ref{algo:GSE}.}

\responseThree{Algorithm \ref{algo:GSE} eliminates $(\ns-\ks)$ sensor locations so that we are effectively left with $\ks$ sensors. The basic idea behind a greedy approach is to make locally optimal decisions with aim to achieve global optimality. Therefore, Algorithm \ref{algo:GSE} begins with the set of all sensors $\Sens$, and iteratively eliminates a sensor such that the optimal cost (i.e., $l_1$-norm of the precision vector) in \eqn{hinf_thm_obs} of the resulting set (after elimination) is minimum. Such iterations are performed  $(\ns-\ks)$ times unless \eqn{hinf_thm_obs} becomes infeasible.}

\responseThree{Since greedy algorithms are only locally optimal, in most cases the resulting solution will be sub-optimal. However, if the cost function satisfies certain special conditions such as modularity or sub/super-modularity, then certain guarantees about the optimality of greedy approaches can be provided \cite{nemhauser_analysis_1978}.
Unfortunately, the cost function under consideration 
can be shown not to exhibit sub/super-modular structure \cite{deshpande_ADMMGreedy_TAC2021}, thus, the well-known theoretical guarantees about the performance of greedy methods can not be used. Despite the lack of special properties of the cost function, empirical results in previous works \cite{zhang_sensor_2017,deshpande_ADMMGreedy_TAC2021} show that the the greedy sensor selection algorithms perform reasonably well in practice. The numerical results in \cite{deshpande_ADMMGreedy_TAC2021} also show that the greedy algorithm performs better than other heuristics in the existing literature.}
\begin{remark}
\responseThree{The optimization problem \eqn{hinf_thm_obs} is solved $|\set{Q}|$ times in each iteration of Algorithm \ref{algo:GSE}. Therefore, the algorithm solves \eqn{hinf_thm_obs} a total of $\frac{\ns(\ns+1)}{2}-\frac{\ks(\ks+1)}{2}$ times to obtain a feasible solution. 
Parallelizing multiple optimization problems in each iteration of the algorithm can significantly reduce overall the execution time.}
\end{remark}
\section{Numerical Results} \label{sec:res}
We solve the sensor placement problem for different values of $M$ and $m$ as discussed below. The optimization problem \eqn{hinf_thm_obs} is solved using the solver \texttt{MOSEK} \cite{mosek_aps_mosek_nodate} with  \texttt{CVX} \cite{grant_cvx_2020} as the parser.

\subsection{Case 1: 10 cells, 1 sensor}
We consider the scenario discussed in \cite{lin_robust_2020}, i.e. there are 10 cells in the string and we are tasked to place one sensor, i.e., $M=10$ and $m=1$. The problem was solved for $S_d = 10$ and the $\Hinf$ performance index $\gamma = 1$, i.e. we require that $\norm{\GO}{\Hinf}<1$. Note that the value $S_d = 10$ could also be interpreted as the maximum possible disturbance of $\pm 10 K$ in the inlet temperature of coolant, $T_{f,\text{in}}$.

\responseOne{The sensor's location and precision are determined using Algorithm \ref{algo:GSE}, i.e., we begin with all possible 10 sensor locations and iteratively eliminate one sensor location at a time until there is only one location left (similar to \fig{case2_m}). According to the output of Algorithm \ref{algo:GSE}, the sensor should be placed at Cell 3 with precision $p=19.99$.} The corresponding scaling parameter $\scln$ is given by $\scln = 1/\sqrt{p} = 0.22$. Therefore, for the desired performance of $\gamma = 1$, the sensor noise power should be $0.22 K$ or sensor measurement error should be within the bounds $\pm 0.22 K$.

We also solved the sensor placement problem via exhaustive search, i.e., the problem \eqn{hinf_thm_obs} was solved ten times with every possible sensor location, and the sensor at Cell 4 was found to require the least precision. However, the difference in optimal cost (sensor precision) as determined by the greedy algorithm and the exhaustive search was found to be of the order of $10^{-8}$. Thus the greedy algorithm was able to recover the optimal cost.


\begin{figure}
    \centering
    \includegraphics[trim={0.08cm 0.05cm 0.08cm 0cm},clip,width=0.46\textwidth]{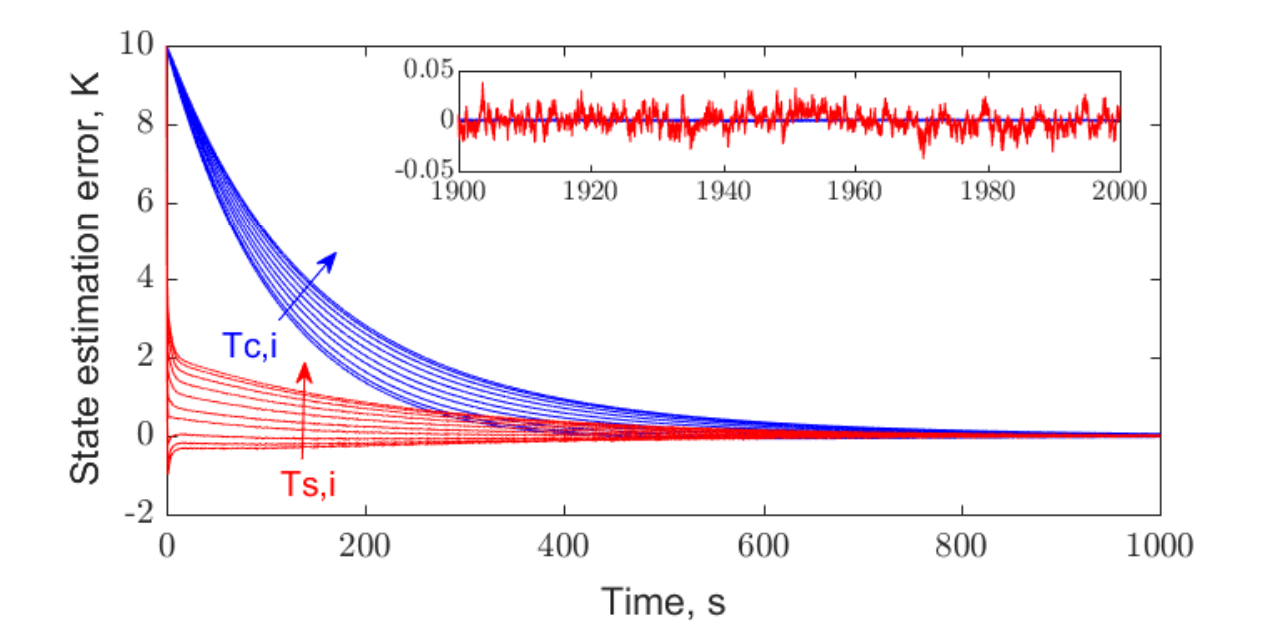}
    \caption{Case 1: Evolution of estimation errors (sensor is placed at Cell 3)}
    \label{fig:case1_err}
    \vspace{-12pt}
\end{figure}
\begin{figure}
    \centering
    \includegraphics[trim={0.1cm 0.05cm 0.1cm 0cm},clip,width=0.46\textwidth]{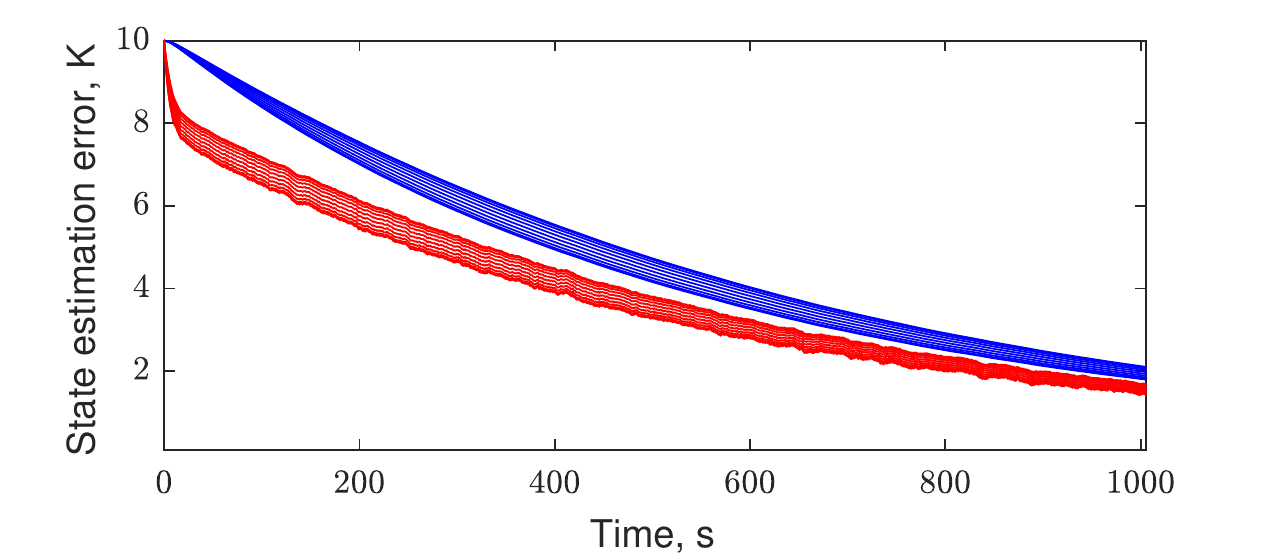}
    \caption{Case 1: Evolution of estimation errors (open loop)}
    \label{fig:case1_err_Op}
    \vspace{-12pt}
\end{figure}
The time evolution of estimation errors is shown in \fig{case1_err} where the error in the initial condition of the thermal field is assumed to be $10 K$ uniform for all states. Blue and red lines, respectively, show the errors in core and surface temperatures, and the cell index increases in the direction of the arrows. 
The observer gain was found to be $\Lg = \Vec{\vo{\tilde{L}}}$ where \responseFive{$\Vec{\cdot}$ denotes the matrix vectorization operator, and}
\begin{align*}
    \vo{\tilde{L}} = -\begin{bmatrix}
    0.0180 &   0.0139 &    0.0108 &    0.0093 &    0.0070\\
    2.8536 &   2.8352 &   2.4741  &  1.8783   & 1.6503\\
    0.0165 &   0.0126 &   0.0105  &  0.0081   & 0.0062\\
    2.6752 &   1.9306 &   1.9728  &  1.7566   & 1.5745
    \end{bmatrix}.
\end{align*}
 From inset of \fig{case1_err}, we observe that estimation errors are bounded within $\pm 0.05K$ asymptotically.

\responseOne{For comparison purpose, the open loop estimation errors are shown in \fig{case1_err_Op} that correspond to $\Lg=\vo{0}$ in \eqn{obs_err}. The $\Hinf$ norm of the open loop error system is $44.72$, significantly greater than the desired bound $\gamma = 1$, which is also apparent from  \fig{case1_err_Op} as errors take much longer to approach zero. }



\subsection{Case 2: 40 cells, 4 sensors}
\begin{figure}
    \centering
    \includegraphics[trim={1cm 0.1cm 1.2cm 0cm},clip,width=0.48\textwidth]{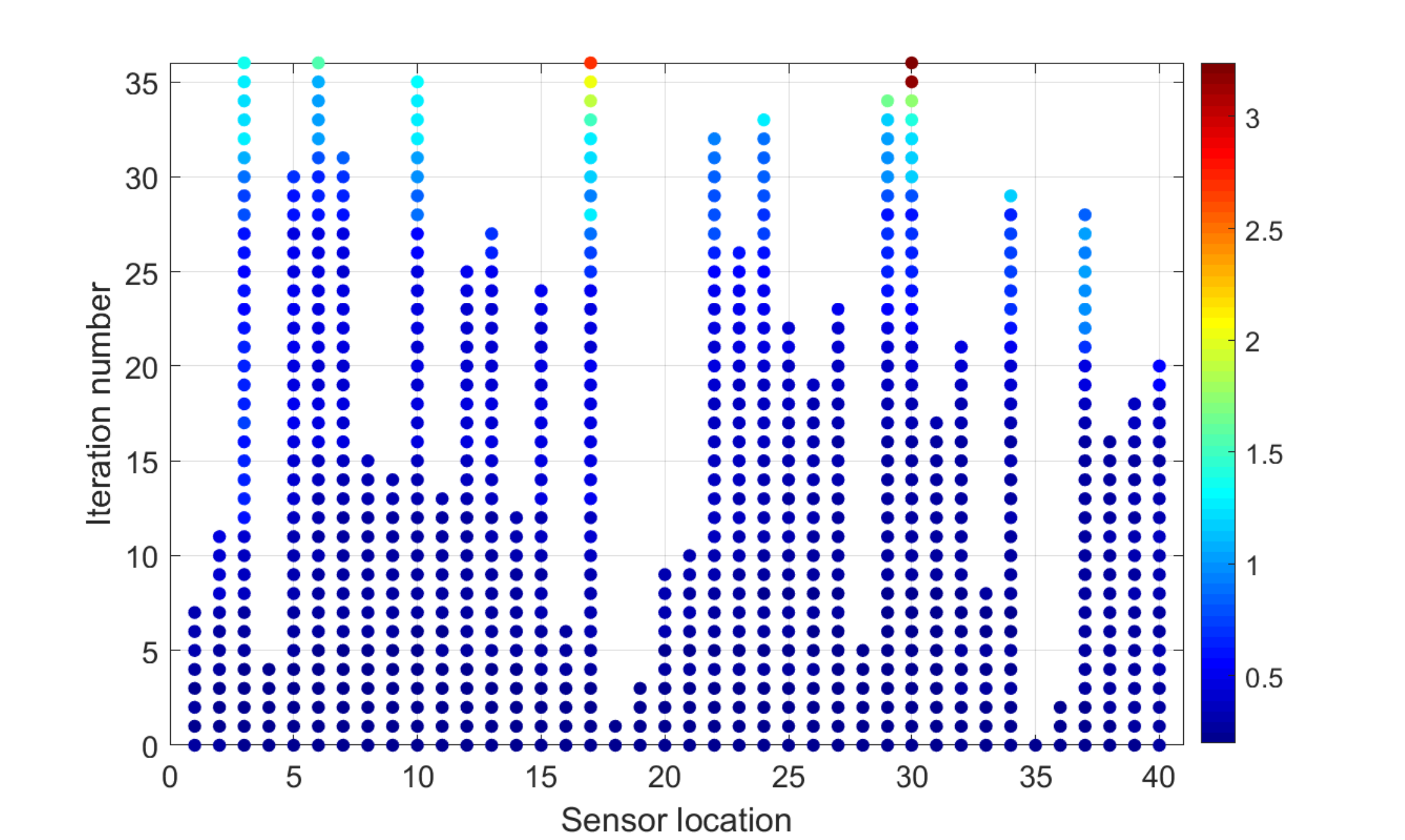}
    \caption{Case 2: Selected sensor locations as iterations of the greedy algorithm progress. Colorbar indicates the corresponding precision values.}
    \label{fig:case2_m}
    \vspace{-10pt}
\end{figure}
Since $M=40$ and $m=4$ for this case, Algorithm \ref{algo:GSE} finds a feasible solution by solving the optimization problem $(40\times 41 \! - \! 4\times 5)/2 \!=\! 810$ times. On the other hand, an exhaustive search to obtain the optimal solution would require the problem to be solved $40!/4!36! = 91390$ times, which is a considerably larger number. Thus, the greedy method in Algorithm \ref{algo:GSE} provides a tractable approach for sensor placement.

\fig{case2_m} shows the selected sensors and their precisions after each iteration of Algorithm \ref{algo:GSE}. Solid circles show the sensors which are not eliminated yet, and their color indicates the corresponding precision as per the color bar on the right.
Results are shown for $S_d = 10$ and $\gamma = 3$.

At the initialization, i.e., iteration zero, all sensors are selected, and in each following iteration, one sensor is eliminated until the cardinality constraint $m=4$ is satisfied. We observe that as the selected locations become sparser with each iteration, the individual sensor precisions (required to satisfy the same $\Hinf$ performance bound) increase. As per the solution obtained at the end of 36$^{\text{th}}$ iteration, sensors should be placed at Cells 3, 6, 17, and 30 with precisions as indicated by the color bar in \fig{case2_m}.

\section{Conclusion} \label{sec:concl}
We discussed the problem of thermal sensor placement for battery systems consisting of several cells packed in a string. An observer with bounded temperature estimation errors was designed in an $\Hinf$ framework while simultaneously minimizing the sensor precisions. The numerical results show that the greedy algorithm discussed in the paper provides a tractable approach for sensor placement in large battery systems. The extensions of this work in the future will consider filter design for thermal field estimation and more complex battery models in terms of different geometries and parametric uncertainties.


\bibliographystyle{unsrt}
\bibliography{MyLibrary}
\end{document}

%% file: macros.tex
\newcommand{\vo}[1]{\boldsymbol{#1}}

\newcommand{\A}{\vo{A}}

\newcommand{\x}{\vo{x}}
\newcommand{\y}{\vo{y}}
\newcommand{\z}{\vo{z}}
\newcommand{\w}{\vo{w}}
\newcommand{\n}{\vo{n}}

\newcommand{\Lg}{\vo{L}} 

\newcommand{\nbat}{M}
\newcommand{\nsens}{m}

\newcommand{\ny}{N_y}

\newcommand{\ns}{\nbat}
\newcommand{\ks}{\nsens}

\newcommand{\precs}{p}
\newcommand{\scln}{\sigma}

\newcommand{\Sens}{\mathcal{S}}

\newcommand{\Hinf}{\mathcal{H}_{\infty}}
\newcommand{\norm}[2]{\left\lVert#1\right\rVert_{#2}}

\newcommand{\Bd}{\vo{B}_d}
\newcommand{\Bu}{\vo{B}_u}

\newcommand{\Bw}{\vo{B}_w}

\newcommand{\Cy}{\vo{C}_y}
\newcommand{\Cz}{\vo{C}_z}

\newcommand{\Dw}{\vo{D}_w}

\newcommand{\zerr}{\vo{\varepsilon}}
\newcommand{\xerr}{\vo{e}}

\newcommand{\GO}{\vo{\mathcal{G}}(s)}

\newcommand{\pow}[1]{\text{pow}(#1)}

\newtheorem{remark}{Remark} 

\newcommand{\xdot}{\dot{\vo{x}}}

\newcommand{\set}[1]{\mathcal{#1}}

\newcommand{\I}[1]{\vo{I}_{#1}}

\newcommand{\X}{\vo{X}}

\newcommand{\Y}{\vo{Y}}

\newcommand{\M}{\vo{M}}

\renewcommand{\u}{\vo{u}}

\newcommand{\ith}{$i^{\text{th}}$ }

\newcommand{\Real}{\mathbb R}

\newcommand{\inner}[1]{\left\langle \vo{e}\phi_i\right\rangle}

\newcommand{\eqnlabel}[1]{\label{eqn:#1}}

\newcommand{\eqn}[1]{(\ref{eqn:#1})}

\newcommand{\fig}[1]{Fig. \ref{fig:#1}}

\DeclareMathAlphabet{\mathbfsf}{\encodingdefault}{\sfdefault}{bx}{n}

\newcommand{\domain}[1]{\set{D}}

\newcommand{\diag}{\textbf{diag}}